 \documentclass[final,5p,times,twocolumn]{elsarticle}

\usepackage{amssymb}
\usepackage{amsmath}
\usepackage{color}
\usepackage{url}
\journal{Nuclear Physics B}

\newcommand{\iso}[2]{{\ensuremath{{}^{#2}\mathrm{#1}}}}
\newcommand{\pthn}{\mbox{$^{190}$Pt} } 

\begin{document}

\begin{frontmatter}



\title{A new precision measurement of the $\alpha$-decay half-life of $^{190}$Pt}


\author[zero]{Mih\'aly Braun}
\ead{mbraun@atomki.mta.hu}
\address[zero]{Institute for Nuclear Research of the Hungarian Academy of Sciences,
Bem t\'er 18, Debrecen, 4026 Hungary}
\author[first]{Yordan M. Georgiev}
\ead{y.georgiev@hzdr.de}

\author[first]{Tommy Sch\"onherr}
\ead{	t.schoenherr@hzdr.de}
\address[first]{Institute of Ion Beam Physics and Materials Research, Helmholtz-Zentrum Dresden-Rossendorf, 01328 Dresden, Germany}
\author[second]{Heinrich Wilsenach \corref{cor1}}
\ead{heinrich.wilsenach@tu-dresden.de}\cortext[cor1]{Corresponding
author}

\author[second]{Kai Zuber}
\ead{zuber@physik.tu-dresden.de}
\address[second]{Institut f\"ur Kern- und Teilchenphysik, Technische Universit\"at Dresden,01069 Dresden, Germany}

%\author{}

%\address{}

\begin{abstract}
A laboratory measurement of the $\alpha$-decay half-life of \pthn has been performed using
a low background Frisch grid ionisation chamber. A total amount of 216.60(17)~mg of natural platinum has
been measured for 75.9~days. The resulting half-life is $(4.97\pm0.16)\times 10^{11}$~years, with
a total uncertainty of  3.2\%. This number is in good agreement with the half-life obtained using the geological comparison method. % \textcolor{blue}{and the most precise laboratory measurement performed}.  
\end{abstract}

\begin{keyword}
Alpha decay; Natural radioactivity; $^{190}$Pt isotope; Pt-Os dating system;  

\PACS 23.60.+e; 29.40.Cs; 27.80.+w; 

\end{keyword}

\end{frontmatter}

 %\linenumbers

\section{Introduction}\label{intro}

\label{intro}
The study of $\alpha$-decays has been important for the understanding of nuclei and their properties for more than a century. Currently these studies still have impact in various areas of nuclear physics, providing information which is valuable and often not accessible otherwise. 
The relation between the half-life and the Q-value of the transition has been established long ago as the Geiger-Nuttall law \cite{gei11}. The Geiger-Nuttall law is generally accepted and has been improved by the
Viola-Seaborg parametrisation \cite{viola2016}. Furthermore, empirical relations were established for a description 
beyond the Geiger-Nuttall law \cite{den09, den09a}. Providing more accurate input data for these studies 
is always helpful.\\
Another important application is the usage of \pthn as cosmochronometer as suggested by \cite{wal97}. Studying terrestrial samples well known from U-Pb dating a value has been  derived  for the decay constant of $^{190}$Pt  of $1.542 \times 10^{-12}$ per year , later corrected by \cite{Begemann} with an error of 1\% to be $1.477 \times 10^{-12}$~per~year. Newer measurements in 2004 \cite{wal04} on iron meteorites 
came to a result of $1.415 \times 10^{-12}$~per~year with a similar uncertainty. Hence both results are not in  very good  agreement with each other  and might indicate a systematic error of about 5\%. Hence an independent laboratory measurement of the $^{190}$Pt half-life of less than 5\% would
shed new light on the obtained values. \\
The $^{190}$Pt-$^{186}$Os system is a powerful tool to investigate the chronology and evolution of geo- and cosmo-chemistry samples. For example, it can be used to understand the history of the abundances of highly siderophile elements in the Earth's mantle. By exploiting the fact that platinum can separate from osmium during partial melting in the mantle, \cite{Brandon} derived that the abundances of these elements were contributed by a late veneer. Furthermore, the age of iron meteorites can also be derived thanks to the different distribution coefficients of Pt and Os during crystallisation \cite{wal04}.\\

Three different methods have been used over almost 90 years to derive a half-life determination of \pthn :
First of all laboratory experiments based on ionisation chambers and nuclear emulsions using the energy-range
relation for $\alpha$-particles. Secondly, geochemical samples from the aforementioned dating of the early solar
system and earth and finally semi-empirical calculations. All the experimental results so for have been compiled in \cite{tav06}. In this reference weighted averages are given of $3.9(2) \times 10^{11}$ years (laboratory measurements), $4.78(5)\times 10^{11}$ years (geological measurements) and $3.5(3) \times 10^{11}$ years 
(semi-empirical approach) respectively. While laboratory and semi-empirical data show more than 10\% uncertainty,
even so individual results show a significantly wider spread, the claimed geological uncertainties are on the level
of 1\% and below.  Furthermore there is a significant difference among the weighted averages of all three
values, so it is
not really clear what the half-life of \pthn is. In addition the recommended value by the Nuclear Data sheets is $6.5(3)\times10^{11}$ years  \cite{nds2003}.\\

The aim of this paper is to provide new data about the half-life of the $^{190}$Pt-decay by performing a laboratory measurement with a precision significantly below 10\%. For that purpose a low background ionisation 
chamber has been used. The Q-value of the decay is 3252(6) keV \cite{ame12}, this corresponds to a peak in the $\alpha$-spectrum at $E_{\alpha}$ = 3183(6)~keV.

\section{Experimental Setup}

The $\alpha$-decay of \iso{Pt}{190} was measured with the use of a twin Frisch-grid ionisation chamber (TFGIC), with a common anode in the middle, an upper chamber usable as a veto and monitor and a lower
chamber for the real measurement. Its diameter is 30 cm and all samples are placed on the bottom. The chamber was  specifically designed for the detection of low rate $\alpha$-decays. This was achieved through constructing a TFGIC with a low level of internal radioactive impurities and in
 adding pulse shape discrimination to separate background from signal. In this way FADC pulses
 were taken from
 the grid and from the anode. This allowed to identify cuts and reject non-physical events.
 As chamber gas P10 was used. The chamber has a background of only 22 counts in the energy region of 1~MeV to 3.5~MeV within a 30.8~day run period. The majority of the events were above 5 MeV as the events originate from Rn-decay products due to the sample changing process and a tiny leak.  
 The design of the chamber and its performance is discussed in more detail in \cite{cham}.

\section{Measurement}

Natural platinum was deposited on three 4" silicon wafers by electron beam evaporation in a LAB~500 evaporation system (Leybold Optics GmbH). The following parameters were used during deposition: 1.3~kW beam power, 0.6~nm/s deposition rate, 300~mm distance between the crucible and the wafer, and $8\times10^{-5}$~Pa initial chamber pressure. To optimise the detection of the $\alpha$-decay from $^{190}$Pt, the coating was produced in the range of 400 nm. The wafers were weighed before and after deposition to determine the amount of platinum deposited. Assuming a density of 21.45(5)~g/cm$^3$, it is possible to estimate the thickness of the deposited layers. This is shown in table~\ref{tab:thickness}. The platinum was obtained from Kurt J. Lesker$^{\circledR}$ and had a purity of 99.99\%.

\begin{table}[htbp]
\caption{Table showing the measured properties of each platinum disc.} 
\label{tab:thickness}
\begin{center}
\begin{tabular}{| l | c c |}
\hline
Target Name & Mass [mg] & Thickness [nm] \\ \hline
Pt01 & 75.2(1) & 441(4) \\ \hline  
Pt02 & 73.1(1) & 429(4) \\ \hline	
Pt03  & 68.3(1) & 401(4) \\ \hline	
\end{tabular}
\end{center}
\end{table}

The three wafers were placed at the bottom of the chamber. They were arranged so that their centres formed a triangle without overlapping. This configuration was chosen so that most of the platinum is close to the centre of the chamber. In this way tiny changes in the detection efficiency close to the side walls can be ignored. Nevertheless, the effect was taken into account in the simulation.

The total measurement time was 75.9 days. The run was started and was only interrupted to renew the counting gas, while the position was not altered and the chamber was not opened. This was done four times to ensure the most stable run conditions. Moreover, this also ensured that the background was reduced as a function of run time, leading to a cleaner run. All of the periods between the flushes were initially treated as separate runs. The rate and energy spectra were then compared to ensure that the run technique was reliable. After this, all of the data was combined. 

The isotopic abundance was measured at Debrecen, Hungary, using inductively coupled plasma mass spectrometry (ICP-MS), the values of which are given in Table~\ref{tab:ab}. The molar mass is calculated from the abundances. This was done by multiplying each abundance with its corresponding mass, and summing the products. The masses of each isotope were taken from \cite{deLaeter2003}. This results in a molar mass of 195.1(11)~g/mole. The abundances were measured and compared with a calibration standard of natural platinum obtained from IRMM (Geel).

\begin{table}[htbp]
\caption{Table is isotopic abundances measured for the used platinum samples. The literature values are taken from \cite{deLaeter2003}.}
\label{tab:ab}
\begin{center}
\begin{tabular}{|r|c|c|}
\hline
 Nucleon  & Abundance & Lit. Abundance \\
Number	&   [\%]  &  [\%] \\ \hline
190 & 0.01125(21) &  0.014(1) \\ 
192 & 0.785(26) & 0.782(7)\\ 
194 & 33.23(35) &32.967(99) \\ 
195 & 33.33(13) & 33.832(10)\\ 
196 & 25.20(42) & 25.242(41)\\ 
198 & 7.4625(90) & 7.163(55) \\ \hline
\end{tabular}
\end{center}
\end{table}

In the measurement there was a small contamination into the $^{190}$Pt signal region from the tailing of background $\alpha$-lines at higher energy. This contamination was fitted with an exponential and the rate was found to be 1.0203(89)~c.p.d in the region of interest (ROI). The ROI extends from 1.5~MeV to 3.7~MeV. This was compared with the background rate in the ROI measured in a background run of 37~days. The rate was found to be 1.10(17) c.p.d., which agrees well with the background taken from the fit. 

\begin{figure}[h!]
\includegraphics[width=0.95\columnwidth]{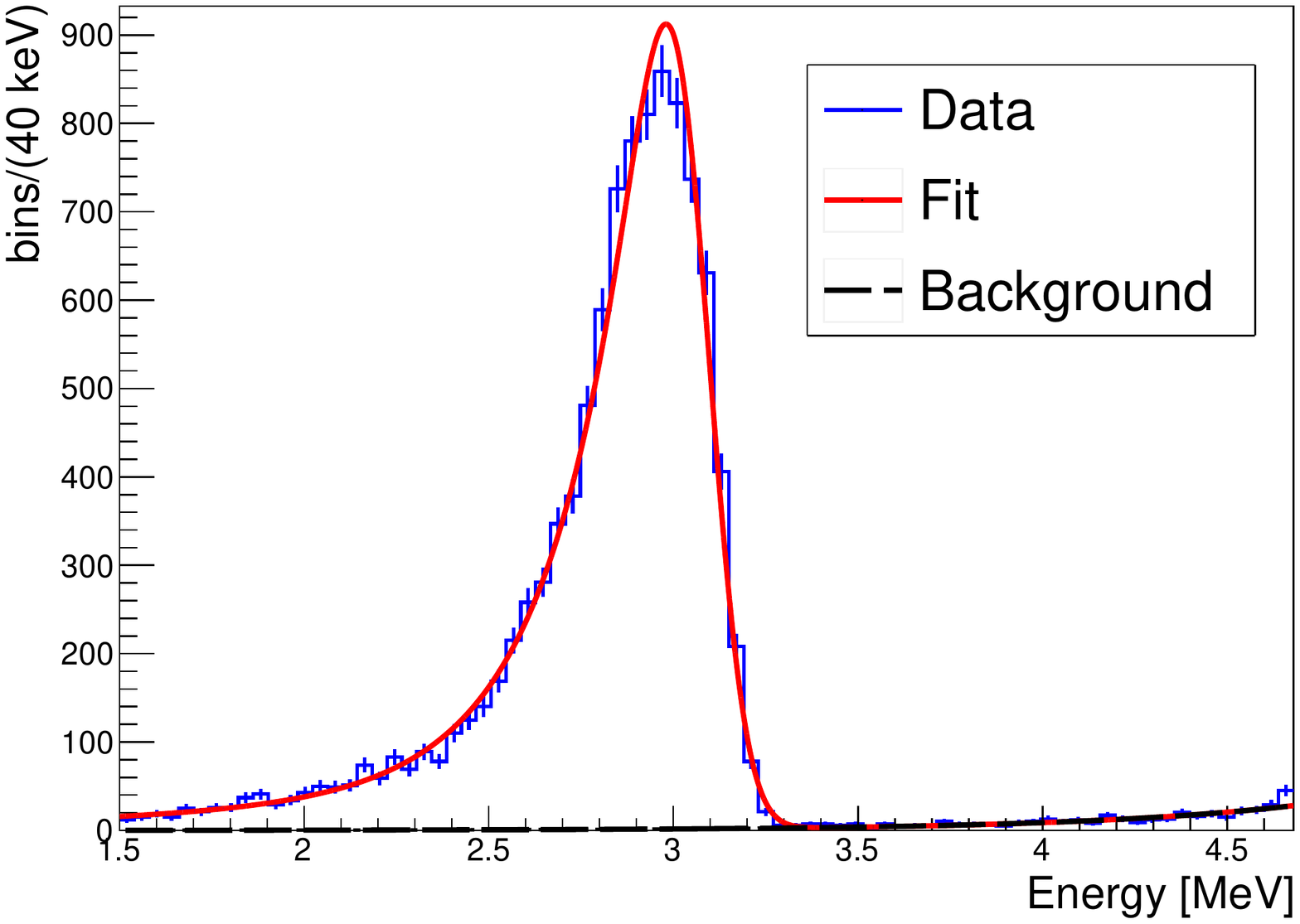}  
\caption{Fitted spectrum of the 75.9 day run. The solid red line is the fit to the data. The dashed black line (only visible on the right side) is the background contamination into the ROI from higher energy. The fit model uses two exponentials which are cut at the peak with an error function. The energy resolution at the peak position is 83.9(2.3)~keV. }
\label{fig1}
\end{figure}

The fit of the energy spectrum around the $^{190}$Pt peak region can be seen in Figure~\ref{fig1}. The signal of the $\alpha$-decay was then fitted with the characteristic $\alpha$ spectrum \cite{fit_func} using the least squared method. The fit gave a reduced $\chi^2$ of 110/71 =  1.55, and the fit parameters were unconstrained. The slight tension in the fit comes from the peak of the distribution. The fitting function assumed that the energy broadening in the spectrum due to self ionisation is less than the energy resolution. For samples of about 400~nm thickness, 
a gaussian peak can be expected as the energy broadening is much smaller than the energy resolution. Though the fit is still a very good approximation. The total number of counts in the ROI was found to be 10180(101). Subtracting the background from this gives a number of signal events of 10103(101), which corresponds to a $^{190}$Pt decay rate of 133.1(13)~c.p.d.   

To obtain the number of \iso{Pt}{190} atoms the samples were weighted. The number of \iso{Pt}{190} atoms is calculated in the following way,

\begin{equation*}
	N_0 = \frac{m}{n} \cdot N_a \cdot a \;,
\end{equation*}

where $m$ is the weighted mass of the platinum, $n$ is the molar mass of the natural platinum, $N_a$ is Avogadro's constant and $a$ is the isotopic abundance of \iso{Pt}{190}. A small number of events are lost due to the inefficiencies of the data cuts and the detection characteristics. This is given as the detection efficiency which was previously measured to be 98.6(22)\% \cite{cham}. A Monte Carlo simulation was performed using the GEANT simulation package. This was used to determine the efficiency of detecting an $\alpha$-particle considering the geometry of the sample. This was found to be 47.014(52)\% as the samples are on the bottom of the chamber. Accounting for these two efficiencies then gives the true rate in the chamber as 287.2(7.0)~c.p.d.. The half-life can then be calculated using the following formula,

\begin{equation*}
	T_{1/2} = \frac{\ln(2)\cdot N_0}{R_t}\;,
\end{equation*}

where $R_t$ is the true signal rate in the chamber. Following this procedure gives a half-life of the ground state transition of;

\begin{equation*}
	T_{1/2} = (4.97 \pm	0.16)\times{10^{11}}~\text{a} \;.
\end{equation*}

The error here is the combination of the statistical and systematic error added in quadrature. The values used to derive the half-life and the sources of uncertainties can be seen in table~\ref{tab:stat}.

\begin{table}[htbp]
\caption{Table showing the sources of uncertainty of each quantity and their relative errors.}
\begin{tabular}{|l c|r|r|}
\hline
 Quantity &	& Value &Relative Error \\ 
 		 & 	&		& [\%] \\	\hline
Detector Eff. &[\%]& 98.6(2.2) & 2.23 \\ 
$^{190}$Pt Abundance  &[\%] & 0.01125(21) & 1.88 \\ 
Signal Rate &[c.p.d]& 133.1(13) & 1.00 \\ 
Molar Mass &[g/mole]& 195.1(11) & 0.57 \\ 
Geometric Eff. &[\%] & 47.014(53) & 0.11 \\ 
Background Rate &[c.p.d] & 1.0203(89) &  0.087  \\ 
Total Mass &[mg] & 216.60(17) & 0.08 \\ \hline
\end{tabular}
\label{tab:stat}
\end{table}

To achieve a ``gold standard" relative uncertainty of 0.2\% (as in the case of $^{238}$U) two very important steps have to be taken. The first is to reduce the uncertainty coming from the sample while still having an adequate activity. This can be achieved for instance by using AMS to select and implant only $^{190}$Pt a few nanometers into an ultra clean target material. The number of implantation isotopes should be characterised to a relative uncertainty below 0.1\%. Enough of the $^{190}$Pt should be implanted to give a reasonable rate in the chamber. To have a statistical uncertainty of around 0.1\%, 1,000,000 events need to be seen in the chamber. A rate of 0.3~Bq could reach the desired statistics in a reasonable time. This would take $6.6\times{10}^{18}$ atoms of $^{190}$Pt. 
The second obstacle would be to eliminate the main systematic uncertainty in the chamber, which is the knowledge of the detector efficiency. This would require a calibration source which has an activity known to the precision of 0.1\%.

\section{Results and Discussions}

In this paper we have presented a new half-life of the $\alpha$-decay of $^{190}$Pt. The aim with respect to previous laboratory measurements was a much smaller error budget. Hence, this is the first half-life measurement of $^{190}$Pt that uses ICP-MS to determine the isotopic abundance. Furthermore the measuring time is about an order of magnitude longer than any other ionisation chamber measurement performed. In this way the counting statistics uncertainty could be reduced to 1\%. Finally, due to 
the design of the chamber being made of low-background materials and the usage of very thin targets  events of the peak tailing towards lower energy are a minor correction. The analysis is based on pulse shape discrimination, that has a high efficiency of discriminating between signal and background. The geometric efficiency has also been determined using a complex simulation.

\begin{figure}[!htb]
\includegraphics[width=0.95\columnwidth]{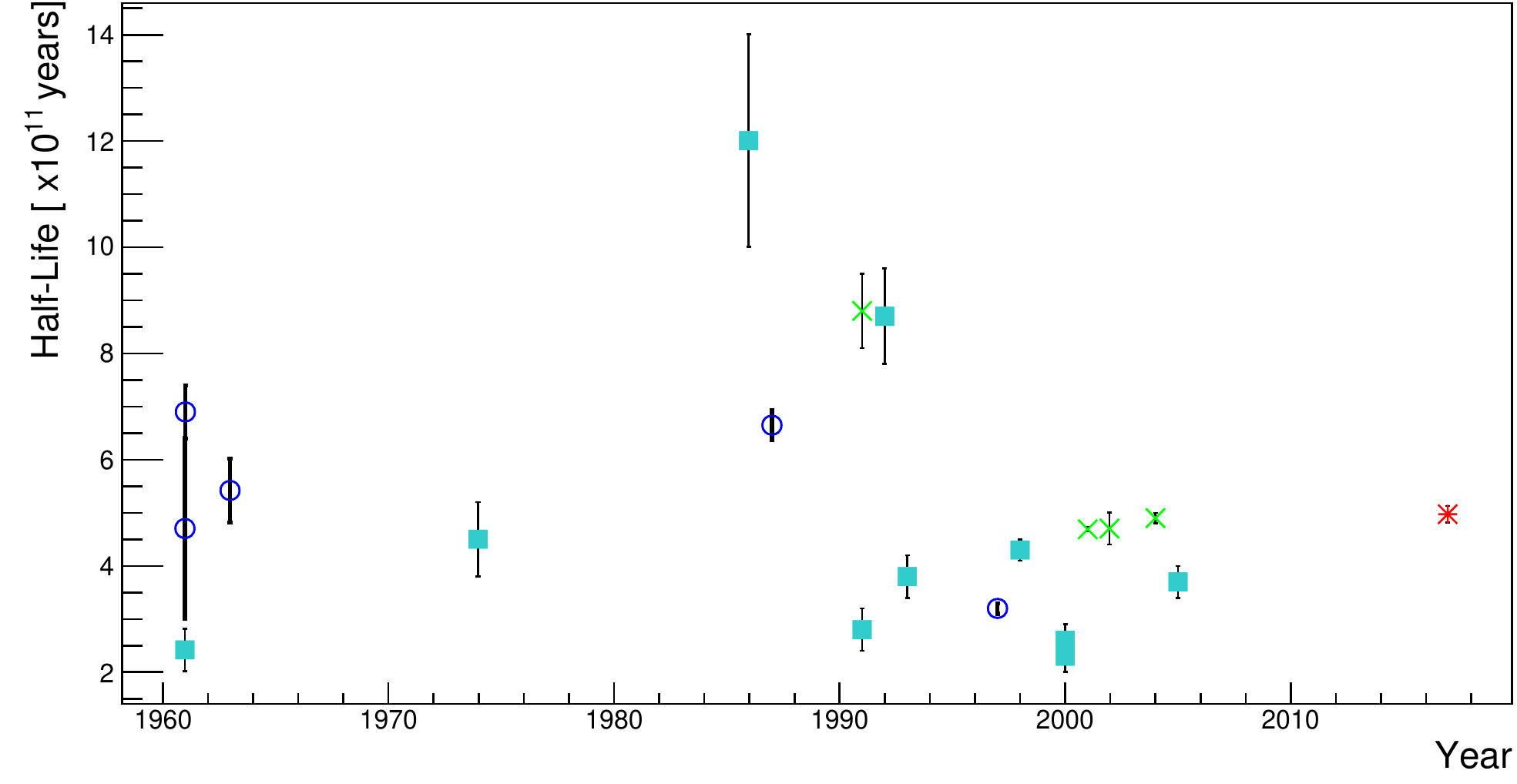}
\caption{\pthn half-life measurements as a function of time. Half-life values are taken from \cite{tav06}.The weighted mean of the of direct counting method (blue circle), 3.9(2)$\times{10^{11}}$~years agrees well with the mean from the semi-empirical calculation model (teal square) 3.5(3)$\times{10^{11}}$~years. These do however not agree well with the mean obtained using the geological comparison method (green cross) 4.78(5)$\times{10^{11}}$~years. The value obtained in this work is shown as a red star. (Note that the value obtained by Walker et al. \cite{wal97} in 1997 was revised by Begemann et al.\cite{Begemann} in 2001, and only the revised value is used.)}
\label{fig:hl}
\end{figure}

The result of 4.97(16)$\times{10^{11}}$~years is in good agreement with the half-life obtained using the geological comparison method of 4.78(5)$\times{10^{11}}$~years. Figure~\ref{fig:hl} shows a graph of the half-life measurements of $^{190}$Pt as a function of year. It is apparent from this graph that the most recent direct counting method results do not agree well with each other, whereas the most recent half-lives determined using the geological comparison method have been consistent with each other. For the first time it has been shown that the geological comparison method is in good agreement with the direct counting method. 

The sources on possible pitfalls of previous direct counting measurements are unclear. It is however apparent that the isotopic abundance was not measured for each half-life determination. It is also interesting to note that the uncertainty from the abundance is not considered. According to ``Atomic weights of the elements: Review 2000" \cite{deLaeter2003} the relative variation between different samples is around 7\%. When this uncertainty is taken into account, the discrepancy becomes smaller, but it is still statistically relevant. Apart from this oversight, the source of the discrepancy is still unclear. Further direct counting measurements could lead to a statistical basis for adjusting or eliminating discrepant data.

\section{Acknowledgements}
\label{thanks}

We would like to thank B. Scheumann from the Institute of Ion Beam Physics and Materials Research, Helmholtz-Zentrum Dresden-Rossendorf for the assistance with the platinum deposition. We would also like to thank M. Lugaro for valuable discussions and information. This work was supported by BMBF (02NUK013B).



\bibliographystyle{elsarticle-num} 
\bibliography{pt190.bib}


%\begin{thebibliography}{00}


%\bibitem{}

%\end{thebibliography}
\end{document}